\documentclass[aps,prc,letterpaper,11pt,twoside,tightenlines,nofootinbib,showpacs,preprint]{revtex4}
\usepackage[latin1]{inputenc}
\usepackage{amsmath}
\usepackage{amsfonts}
\usepackage{amssymb}
\usepackage{graphicx}
\usepackage{subfigure}
\usepackage{rotating}
\begin{document}

\title{A case in favor of the $N^*(1700)(3/2^-)$.}

\author{E. J. Garzon$^1$, J. J. Xie$^{2,3}$ and E. Oset$^1$}
\affiliation{$^1$ Departamento de F\'{\i}sica Te\'orica and IFIC, Centro Mixto Universidad de Valencia-CSIC,
Institutos de Investigaci\'on de Paterna, Aptdo. 22085, 46071 Valencia,Spain}
\affiliation{$^2$Institute of Modern Physics, Chinese Academy of Sciences, Lanzhou 730000, China}
\affiliation{$^3$State Key Laboratory of Theoretical Physics, Institute of Theoretical Physics, Chinese Academy of Sciences, Bejing 100190,
China}

\date{\today}

\begin{abstract} 
Using an interaction extracted from the local hidden gauge Lagrangians, which brings together vector and pseudoscalar mesons, and the coupled channels $\rho N$ (s-wave), $\pi N$ (d-wave), $\pi \Delta$ (s-wave) and $\pi \Delta$ (d-wave), we look in the region of $\sqrt s =1400-1850$ MeV and we find two resonances dynamically generated by the interaction of these channels, which are naturally associated to the $N^*(1520) (3/2^-)$ and $N^*(1700) (3/2^-)$. The $N^*(1700) (3/2^-)$ appears neatly as a pole in the complex plane. The free parameters of the theory are chosen to fit the $\pi N$ (d-wave) data. Both the real and imaginary parts of the $\pi N$ amplitude vanish in our approach in the vicinity of this resonance, similarly to what happens in experimental determinations, what makes this signal very weak in this channel. This feature could explain why this resonance does not show up in some experimental analyses, but the situation is analogous to that of the $f_0(980)$ resonance, the second scalar meson after the $\sigma (f_0(500))$ in the $\pi \pi$(d-wave) amplitude.
The unitary coupled channel approach followed here, in connection with the experimental data, leads automatically to a pole in the 1700 MeV region and makes this second $3/2^-$ resonance unavoidable.
\end{abstract}
\pacs{11.80.Gw, 12.38.Gc, 12.39.Fe, 13.75.Lb}

\maketitle

\section{Introduction}
\label{Intro}

The resonance $D_{13}(1520)(3/2^-)$ is catalogued as a four star resonance in the PDG
\cite{pdg}. The $D_{13}(1700)(3/2^-)$ is also catalogued as a three star resonance and has been advocated by many groups, the latest ones \cite{Manley:1992yb,batinic,thoma}. However, the $D_{13}(1700)(3/2^-)$ is not ``seen'' in the GWU analysis of \cite{arndt06} nor in former analyses of this group \cite{arndt04}. In a recent unified Chew-Mandelstam SAID analysis of pion photoproduction data the resonance is not needed again \cite{workman}, and it is also not included in the MAID analysis of photoproduction data \cite{maid} where only four star resonances are considered, but it is one of the resonances included in the analysis of the Bonn-Gatchina group \cite{thoma} and shows more clearly in the analysis of the $(\gamma, \pi^0 \pi^0)$ data in \cite{thoma2,Thoma:2007bm}. One common conclusion from \cite{Manley:1992yb,thoma2} is the strong coupling of the $D_{13}(1700)(3/2^-)$ state to the $\pi \Delta$ (d-wave) channel, something not intuitive nor expected from ordinary quark models. 
We should note that, as discussed in \cite{Arndt:2009nv} there could be some reason to miss some resonances in the analysis if the resonances are too wide ($\Gamma >$500 MeV) or they posses a small ratio (BR$<4\%$) to the channel under consideration.
  
In view of the current discussion about these two resonances, and particularly the doubts casted on the existence of the $D_{13}(1700)(3/2^-)$, we have done a different analysis, purely theoretical, although paying attention to known data on the $D_{13}$ channel. The work consist on taking four coupled channels, also considered in \cite{Manley:1992yb,arndt04,arndt06}, the  $\rho N$ (s-wave), $\pi N$ (d-wave), $\pi \Delta$ (s-wave) and $\pi \Delta$ (d-wave). We look in the region of $\sqrt s =1500-1750$ MeV and take the interaction between these channels from the local hidden gauge lagrangians \cite{hidden1,hidden2,hidden3,hidden4}, then solve the Bethe Salpeter equation in coupled channels and from this fully unitary approach we find two resonances dynamically generated by the interaction of these channels, which are naturally associated to the $N^*(1520) (3/2^-)$ and $N^*(1700) (3/2^-)$ resonances. 
The freedom of the theory in the choice of subtraction constants to regularize the loops is used to fit the data on the $\pi N$ (d-wave) amplitude.
The study provides the poles and the residues of the resonances and their coupling to the different channels, widths and partial decay widths. A fair agreement is obtained with phenomenology and the two resonances appear as poles in the complex plane. In particular a relatively strong coupling of the $N^*(1700) (3/2^-)$ to the $\pi \Delta$ (d-wave) suggested in \cite{Manley:1992yb,thoma2} is confirmed. 
  
The results that we find are also illustrative on why some analyses do not find a compelling need for the $N^*(1700) (3/2^-)$ resonance and we clarify the situation by comparing the amplitudes obtained in this case with the $\pi \pi$ amplitude in the scalar-isoscalar sector, where the $f_0(500)$ and $f_0(980)$ resonances appear. The presence of a first resonance with a width that makes it overlap with a second one with the same quantum numbers has consequences on the second resonance and the amplitude with these quantum numbers, producing a zero of the $\pi N$ or $\pi \pi$ amplitudes, respectively, in the vicinity of the second resonance, which makes its coupling to these channels very weak and makes difficult the identification of the $N^*(1700) (3/2^-)$ resonance in the $\pi N$ channel.

\section{Formalism}
We follow the formalism of the hidden gauge interaction of vector mesons \cite{hidden2}. For the interaction of the vectors among themselves we have the three vector Lagrangian
\begin{equation}
{\cal L}^{(3V)}_{III}=ig\langle (\partial_\mu V_\nu -\partial_\nu V_\mu) V^\mu V^\nu\rangle
\label{l3V}\ ,
\end{equation}
where $V_\mu$ is the SU(3) matrix for the nonet of the $\rho$
\begin{equation}
V_\mu=\left(
\begin{array}{ccc}
\frac{\rho^0}{\sqrt{2}}+\frac{\omega}{\sqrt{2}}&\rho^+& K^{*+}\\
\rho^-& -\frac{\rho^0}{\sqrt{2}}+\frac{\omega}{\sqrt{2}}&K^{*0}\\
K^{*-}& \bar{K}^{*0}&\phi\\
\end{array}
\right)_\mu \ ,
\label{Vmu}
\end{equation}
and $g=\frac{M_V}{2 f}$, with $f$=93 MeV.

In the same way, the coupling of the vectors to pseudoscalar mesons is given by 
\begin{equation}
{\cal L}_{VPP}= -ig \langle [
P,\partial_{\nu}P]V^{\nu}\rangle \ ,
\label{lagrVpp}
\end{equation}
where here $P$ is the SU(3) matrix of the pseudoscalar mesons, 
\begin{equation}
P=\left(
\begin{array}{ccc}
\frac{\pi^0}{\sqrt{2}}+\frac{\eta_8}{\sqrt{6}}&\pi^+& K^{+}\\
\pi^-& -\frac{\pi^0}{\sqrt{2}}+\frac{\eta_8}{\sqrt{6}}&K^{0}\\
K^{-}& \bar{K}^{0}&-\frac{2}{\sqrt{6}} \eta_8\\
\end{array}
\right) \ .
\label{Pmatrix}
\end{equation}

The lagrangian for the coupling of the vector to the baryon is given by
\begin{equation}
{\cal L}_{BBV} =
g\left( \langle \bar{B}\gamma_{\mu}[V^{\mu},B]\rangle + 
\langle \bar{B}\gamma_{\mu}B \rangle \langle V^{\mu}\rangle \right) \ ,
\label{lbbv}
\end{equation}
where $B$ is now the SU(3) matrix of the baryon octet
\begin{equation}
B =
\left(
\begin{array}{ccc}
\frac{1}{\sqrt{2}} \Sigma^0 + \frac{1}{\sqrt{6}} \Lambda &
\Sigma^+ & p \\
\Sigma^- & - \frac{1}{\sqrt{2}} \Sigma^0 + \frac{1}{\sqrt{6}} \Lambda & n \\
\Xi^- & \Xi^0 & - \frac{2}{\sqrt{6}} \Lambda
\end{array}
\right) \ .
\end{equation}
For the transitions $V B \rightarrow P B$ we need the VPP Lagrangian of Eq.~(\ref{lagrVpp}) and one of the two pseudoscalar mesons is exchanged between the external VP and the baryon. The coupling of the pseudoscalar to the baryon for the two SU(3) octets is given by
\begin{equation}
{\cal L}_{BBP} =
\frac{F}{2} \langle \bar{B}\gamma_{\mu} \gamma_5 [u^{\mu},B]\rangle + 
\frac{D}{2} \langle \bar{B}\gamma_{\mu} \gamma_5 \left\lbrace u^{\mu}, B \right\rbrace \rangle
\label{lbbp}
\end{equation}
where $F=0.51$, $D=0.75$. A generalization for the case of $\pi N \Delta$ is given below.

In a nonrelativistic approximation that we follow, Eq.~(\ref{lbbp}) provides a vertex
\begin{equation}
-i t_{\pi^0 p p} = \frac{F+D}{2 f_\pi} \vec{\sigma} \cdot \vec{q}
\label{pinn}
\end{equation}
where $\vec{q}$ is the incoming momentum of the $\pi^0$, and other charge combinations are trivially derived using isospin symmetry. In practice one substitutes $(F+D)/2 f_\pi$ by $f_{\pi N N}/m_\pi$ where $f_{\pi N N}=0.935$, empirically determined. Similarly for the $\pi N \Delta$ transition we take 
\begin{equation}
-i t_{\pi^+ p \Delta^{++}} = (-) \frac{f_{\pi N \Delta}}{m_\pi} \vec{S} \cdot \vec{q}
\label{pindelta}
\end{equation}
where the minus sign stems from the phase convention $|\pi^+\rangle = - |1,1\rangle$ of isospin, with an empirically value of $f_{\pi N \Delta}=2.23$. The operator $\vec{S}^+$ is the transition spin operator from spin $1/2$ to $3/2$ normalized such that
\begin{equation}
\left\langle 3/2 M | S^+_\nu | 1/2 m \right\rangle = {\cal C}(1/2,1,3/2;m,\nu,M)
\label{snorm}
\end{equation}
with $S^+_\nu$ written in spherical basis and ${\cal C}(1/2,1,3/2;m,\nu,M)$ the Clebsch-Gordan coefficient.

As shown in Ref.~\cite{Oset:2009vf}, the leading term of the $V B \rightarrow V B$ interaction involves the three vector vertex of Eq.~(\ref{l3V}), with one vector meson exchanged, and the coupling of this exchanged vector to the baryon, given by Eq.~(\ref{lbbv}). The potential provided by this term, keeping the dominant $\gamma ^0$ term in Eq.~(\ref{lbbv}), is given by
\begin{equation}
V_{i j}= - C_{i j} \, \frac{1}{4 f^2} \, \left( k^0 + k^\prime{}^0\right)~\vec{\epsilon}\,\vec{\epsilon }\,^\prime
\label{kernel}
\end{equation}
where $k^0, k^\prime{}^0$ are the energies of the incoming and outgoing vector mesons.
The result of Eq.~(\ref{kernel}) with the $\vec{\epsilon}\,\vec{\epsilon }\,^\prime$ factor for the polarization of the vector mesons stems from considering the three momentum of the external vectors small with respect to the mass of the vector mesons \cite{Oset:2009vf}.
The $C_{i j}$ coefficients can be found in Appendix A of Ref.~\cite{Oset:2009vf}, where the subindex $i$ and $j$ correspond to the different channels for all the states of isospin and strangeness.

This potential has been used as the input of the Bethe-Salpeter equation to study the scattering matrix, 
\begin{equation}
T = [1 - V \, G]^{-1}\, V
\label{eq:Bethe}
\end{equation}
where G is the loop function of a vector meson and a baryon which is calculated in dimensional regularization, as shown in Ref.~\cite{Oller:2000fj,Oset:2001cn}, is given by
\begin{eqnarray}
G_{l} &=& i 2 M_l \int \frac{d^4 q}{(2 \pi)^4} \,
\frac{1}{(P-q)^2 - M_l^2 + i \epsilon} \, \frac{1}{q^2 - m^2_l + i
\epsilon}  \nonumber \\ &=& \frac{2 M_l}{16 \pi^2} \left\{ a_l(\mu) + \ln
\frac{M_l^2}{\mu^2} + \frac{m_l^2-M_l^2 + s}{2s} \ln \frac{m_l^2}{M_l^2} +
\right. \nonumber \\ 
& &  +
\frac{\bar{q}_l}{\sqrt{s}}
\left[
\ln(s-(M_l^2-m_l^2)+2\bar{q}_l\sqrt{s})+
\ln(s+(M_l^2-m_l^2)+2\bar{q}_l\sqrt{s}) \right. \nonumber  \\
& & \left. \left. - \ln(-s+(M_l^2-m_l^2)+2\bar{q}_l\sqrt{s})-
\ln(-s-(M_l^2-m_l^2)+2\bar{q}_l\sqrt{s}) \right]
\right\} \ ,
\label{eq:gpropdr}
\end{eqnarray}
with $\mu$=800 MeV a regularization scale and $\alpha_l(\mu)$ the subtraction constant, depending on the channel.

In the cases where the iteration of the Bethe-Salpeter equation includes the $\rho$ meson or $\Delta$, which have relatively large widths, a convolution of the loop function $G$ with the mass distribution is needed. So the loop function with the convolution for the case of the $\rho$ meson would be
\begin{eqnarray}
\tilde{G}(s)= \frac{1}{N}\int^{(m_\rho+2\Gamma_\rho)^2}_{(m_\rho-2\Gamma_\rho)^2}d\tilde{m}^2
\left(-\frac{1}{\pi}\right) 
{\rm Im}\,\frac{1}{\tilde{m}^2-m_\rho^2+{\rm i} \tilde{m} \Gamma(\tilde{m})}
& G(s,\tilde{m}^2,M^2_B)\ ,
\label{Gconvolution}
\end{eqnarray}
where $\tilde{G}$ is normalized with
\begin{equation}
N=\int^{(m_\rho+2\Gamma_\rho)^2}_{(m_\rho-2\Gamma_\rho)^2}d\tilde{m}^2
\left(-\frac{1}{\pi}\right){\rm Im}\,\frac{1}{\tilde{m}^2-m^2_\rho+{\rm i} \tilde{m} \Gamma(\tilde{m})} \ .
\label{Norm}
\end{equation}
Considering the width of the $\rho$, $\Gamma_\rho$=149.4 MeV, the $\Gamma(\tilde{m})$ function is energy dependent and is given in Ref.~\cite{Geng:2008gx} as
\begin{equation}
\tilde{\Gamma}(\tilde{m})=\Gamma\frac{q^3_\mathrm{off}}{q^3_\mathrm{on}}\theta(\tilde{m}-m_1-m_2)
\end{equation}
with $m_1=m_2=m_\pi$ for the $\rho$ using
\begin{equation}
q_\mathrm{off}=\frac{\lambda^{1/2}(\tilde{m}^2,m_\pi^2,m_\pi^2)}{2\tilde{m}},\quad
q_\mathrm{on} =\frac{\lambda^{1/2}(m_\rho^2,m_\pi^2,m_\pi^2)}{2 m_\rho}
\end{equation}
where $\lambda$ is the K\"allen function and $\Gamma$ is the nominal width of the $\rho$.

Similarly, for the case of the $\Delta$:
\begin{eqnarray}
\tilde{G}(s)= \frac{1}{N}\int^{M_\Delta+2\Gamma_\Delta}_{M_\Delta-2\Gamma_\Delta}d\tilde{M}
\left(-\frac{1}{\pi}\right) 
{\rm Im} \frac{1}{\tilde{M}-M_\Delta+{\rm i} \frac{\Gamma(\tilde{M})}{2}}
& G(s,m^2,\tilde{M}^2)
\label{Gcondel}
\end{eqnarray}
where $\tilde{G}$ is normalized with
\begin{equation}
N=\int^{M_\Delta+2\Gamma_\Delta}_{M_\Delta-2\Gamma_\Delta}d\tilde{M}
\left(-\frac{1}{\pi}\right){\rm Im} \frac{1}{\tilde{M}-M_\Delta+{\rm i} \frac{\Gamma(\tilde{M})}{2}}
\label{Normdel}
\end{equation}
where for the width of the $\Delta$ we take $\Gamma_\Delta$=120.0 MeV.

Once the scattering matrix is evaluated, some peaks appear that can be associated to states. Next step is to find the poles associated to those peaks, in order to obtain the couplings of these states to the different channels. The method used is to search poles in the second Riemann sheet, changing the momentum $\vec{q}$ to $-\vec{q}$ in the analytical formula of the $G$ function when $Re(\sqrt{s})$ is over the threshold of the corresponding channel. Using this method one can find poles, as $(M_R+i\Gamma /2)$, where the real part correspond to the mass of the resonance and the imaginary part is half of the width of this state. However, the convolution of the $G$ function eventually can make the pole disappear in channels with the $\rho$ meson. In this case one can study the amplitude in the real axis using that near the peak the T matrix will be as
\begin{equation}
T_{i j} = \frac{g_i g_j}{\sqrt{s}-M_R+i \Gamma /2}
\label{eq:polematrix}
\end{equation}
where $M_R$ is the position of the maximum  and $\Gamma$ the width at half-maximum. The constants $g_i$ and $g_j$ are the couplings of the resonances to the channels $i$, $j$. Then one can take the diagonal channel and obtain
\begin{equation}
|g_i|^2 = \frac{\Gamma}{2} \sqrt{|T_{ii}|^2}
\end{equation}
where the coupling $g_i$ has an arbitrary phase. With one coupling determined, we can obtain the other ones from the $T_{ij}$ matrices using Eq.~(\ref{eq:polematrix}), given by
\begin{equation}
g_j=g_i \frac{T_{i j}(\sqrt(s)=M_R)}{T_{i i}(\sqrt(s)=M_R)} \ .
\label{eq:polerealtion}
\end{equation}

Once, we obtain the couplings of the resonances for each channel, we can calculate the partial decay widths using the equation
\begin{equation}
\Gamma_i = \frac{1}{2\pi} \frac{M_B}{M_R} p_i g_i^2
\label{eq:decaywidth}
\end{equation}

In the case of the decay channel $\rho N$, both resonances are under the threshold. So the momentum used in the previous equation should be imaginary. Experimentally the decay of the resonance to $\rho N$ is observed because the width of the $\rho$ is big enough to allow the decay, although the resonance is under the threshold. In order to generate this effect in our calculation of the partial decay width, we make the convolution of the momentum with the mass of the $\rho$.

\begin{equation}
\tilde{p}= \frac{1}{N}\int^{(m_\rho+2\Gamma_\rho)^2}_{(m_\rho-2\Gamma_\rho)^2}d\tilde{m}^2
\left(-\frac{1}{\pi}\right) 
{\rm Im}\,\frac{1}{\tilde{m}^2-m_\rho^2+{\rm i} \tilde{m} \Gamma(\tilde{m})}
\dfrac{\lambda^{1/2}(M_R^2,\tilde{m}^2, M_N^2)}{2 M_R}
\theta(M_R-\tilde{m}_\rho- M_N)
\label{prhocon}
\end{equation}
Here $N$ is the same normalization as used in Eq.~(\ref{Norm}).

\section{Theoretical Approach}

\begin{figure}
\subfigure[$\rho N(s) \rightarrow \rho N(s)$]{
\includegraphics[width=0.2\textwidth]{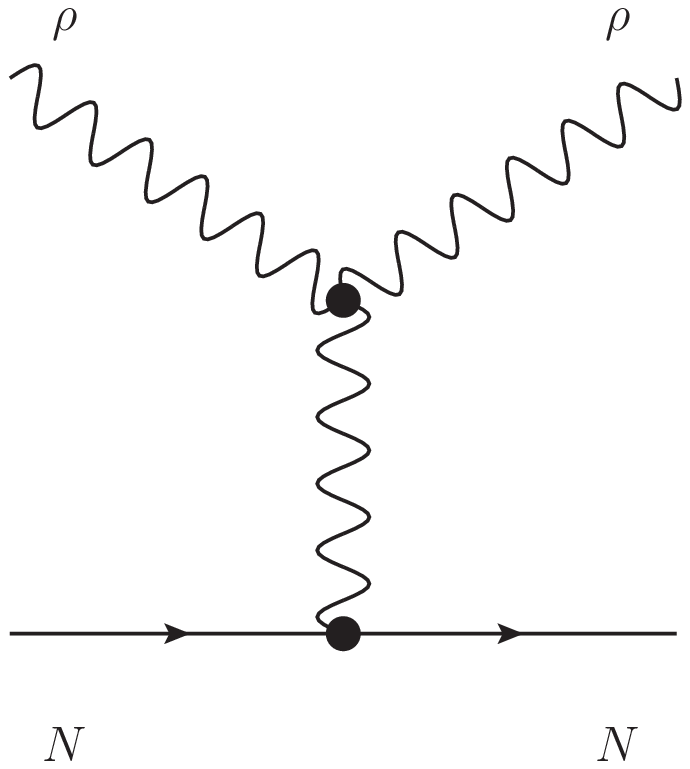}}
\subfigure[$\pi \Delta(s) \rightarrow \pi \Delta(s)$]{
\includegraphics[width=0.2\textwidth]{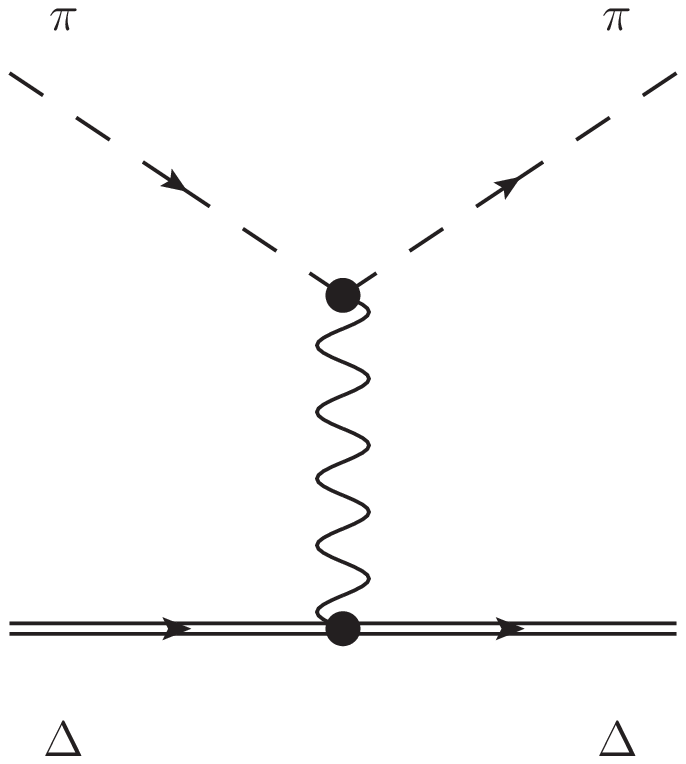}}
\subfigure[$\rho N(s) \rightarrow \pi N(d)$]{
\includegraphics[width=0.2\textwidth]{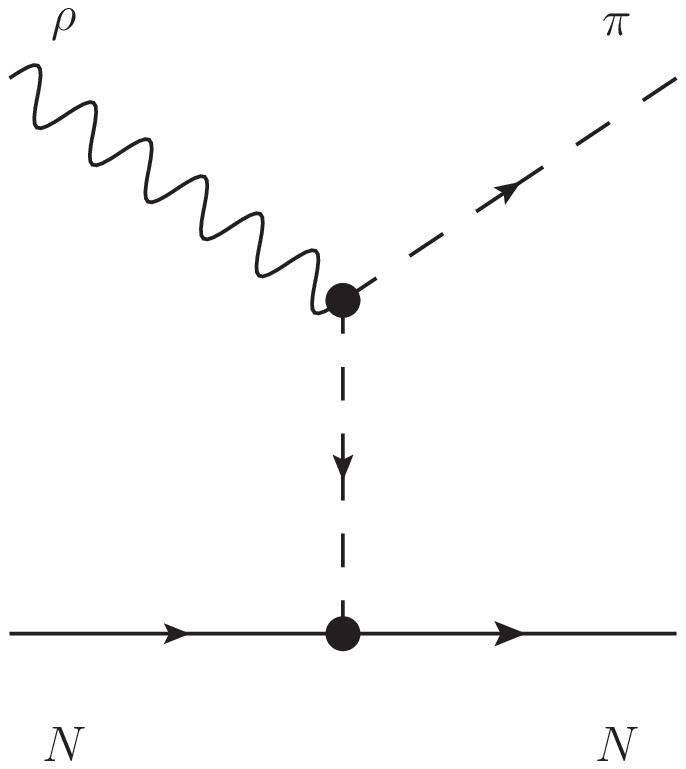}}
\\
\subfigure[$\rho N(s) \rightarrow \pi \Delta(s,d)$]{
\includegraphics[width=0.2\textwidth]{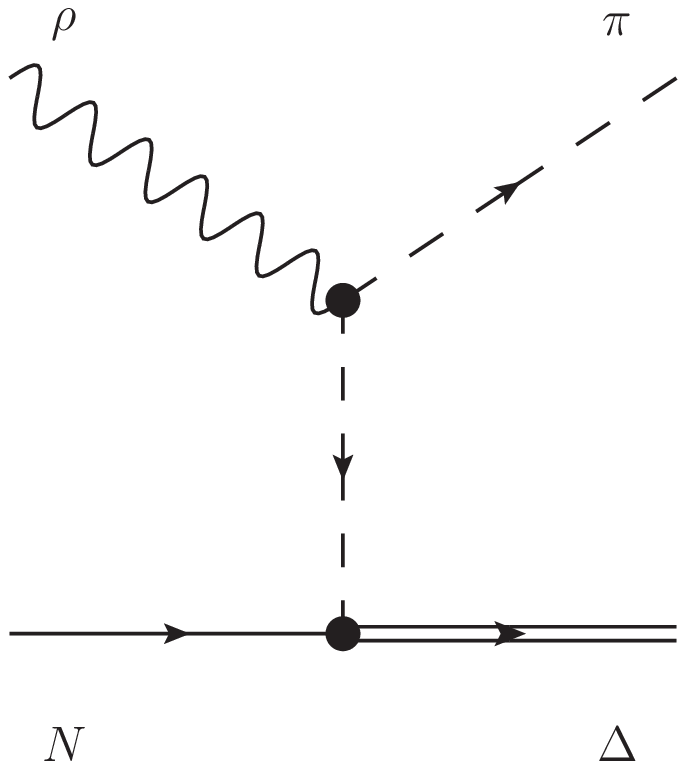}}
\subfigure[$\rho N(s) \rightarrow \pi \Delta(s)$]{
\includegraphics[width=0.2\textwidth]{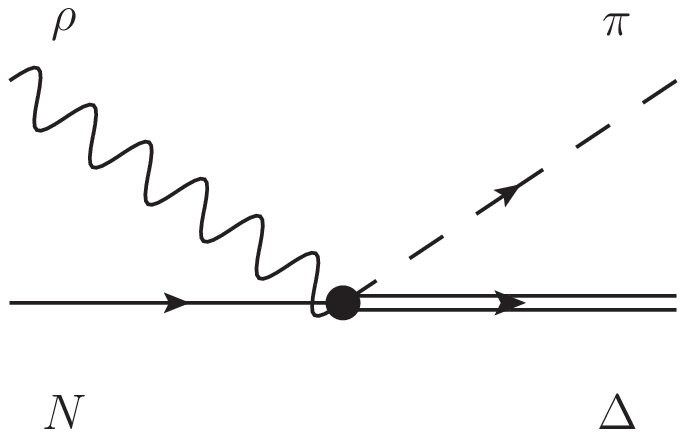}}
\caption{Diagrams of the channels involved in the calculation for $N^*(1520)$ and $N^*(1700)$.}
\label{fig:diagrams}
\end{figure}

As we have commented in the first section, the channels involved in our study are the $\rho N$ (s-wave), $\pi N$ (d-wave), $\pi \Delta$ (s-wave) and $\pi \Delta$ (d-wave) all of them in isospin $I=1/2$. In order to develop our calculation, we need the diagrams of the elastic interaction and the transitions diagrams as well. In Fig. \ref{fig:diagrams} we show those diagrams.
The evaluation of the diagrams shown in Fig. \ref{fig:diagrams} leads us to the following vertices for the transitions (See Appendix \ref{app:vertices} for details)
\begin{equation}
t_{\rho N (s) \rightarrow \rho N (s)} = - \frac{2}{4 f^2} \left( k_{\rho}^0+k_{\rho}^{\prime 0} \right)
\end{equation}
\begin{equation}
t_{\pi \Delta (s) \rightarrow \pi \Delta (s)} = - \frac{5}{4 f^2} \left( k_{\pi}^0+ k_{\pi}^{\prime 0} \right)
\end{equation}
\begin{equation}
t_{\rho N (s) \rightarrow \pi \Delta (s)}=g \frac{2}{\sqrt{3}} \frac{f_{\pi N \Delta}}{m_\pi} \left\lbrace \frac{\frac{2}{3} \vec{q}~^2}{\left(P_V+q\right)^2-m_\pi^2}+1 \right\rbrace
\label{t3}
\end{equation}
\begin{equation}
t_{\rho N (s) \rightarrow \pi \Delta (d)}=g \frac{2}{\sqrt{3}} \frac{f_{\pi N \Delta}}{m_\pi} \left\lbrace \frac{\frac{2}{3} \vec{q}~^2}{\left(P_V+q\right)^2-m_\pi^2}\right\rbrace
\end{equation}
\begin{equation}
t_{\rho N (s) \rightarrow \pi N (d)}=g (-2 \sqrt{6}) \frac{f_{\pi N N}}{m_\pi} \left\lbrace \frac{\frac{2}{3} \vec{q}~^2}{\left(P_V+q\right)^2-m_\pi^2} \right\rbrace
\end{equation}
where we take $f_{\pi N N}=0.935$ and $f_{\pi N \Delta}=2.23$, $k_{\rho}^0$ and $k_{\pi}^0$ are the energies of the $\rho$ and the $\pi$ in the center of mass.

The term of Fig. \ref{fig:diagrams}(e) is the Kroll-Ruderman term which appears in the $\rho N \rightarrow \pi \Delta (s)$ transition. It has the type
\begin{equation}
-i t_{KR} = g \sqrt{2} \frac{f_{\pi N \Delta}}{m\pi} \sqrt{\frac{2}{3}} \vec{S}^+ \cdot \vec{\sigma}
\label{kr}
\end{equation}
and only involves $L=0$, providing the term $1$ in the bracket of Eq.~(\ref{t3}). 

In order to include the $L=2$ transitions, we use the same procedure as used in Ref.~\cite{Roca:2006sz} where the unknown potentials are introduced with a parameter $\gamma$ and the momenta of the L=2 transition. In our case those channels are $\pi \Delta (d) \rightarrow \pi \Delta (d)$, $\pi \Delta (d) \rightarrow \pi N (d)$, $\pi N (d) \rightarrow \pi N (d)$.
With the purpose of working with more suitable parameters $\gamma_{ij}$, we normalized them with the pion mass as following.
\begin{equation}
t_{\pi \Delta (d) \rightarrow \pi \Delta (d)} = -\frac{\gamma_{33}}{m_{\pi}^5} q_3^4
\end{equation}
\begin{equation}
t_{\pi \Delta (d) \rightarrow \pi N (d)} = -\frac{\gamma_{34}}{m_{\pi}^5} q_3^2 q_4^2
\end{equation}
\begin{equation}
t_{\pi N (d) \rightarrow \pi N (d)} = -\frac{\gamma_{44}}{m_{\pi}^5} q_4^4
\end{equation}
where $q_i$ are the momenta of each channel. With this notation $\gamma_{ij}$ have no dimensions and are of the order of 0.01.
Note that the L=2 transition leads us to introduce a $q^4$ term, which at high energies have a very fast grow.
To control this divergence we introduce the Blatt-Weisskopf barrier-penetration factors (See Ref.~\cite{Manley:1984jz}). In the case of $L=2$, we have a substitution as
\begin{equation}
q^2\rightarrow \frac{x^2}{\sqrt{9+3 x^2+x^4}}
\end{equation}
here $x=R q$ where $R=0.25$ fm.

We decided to introduce this factor normalized at the energy of 1700 MeV. Let us call ${x_i}_0=R {q_i}_0$ with ${q_i}_0$ the momentum for a channel $i$ at 1700 MeV. We introduce a factor $B_i$ defined as
\begin{equation}
B_i = \frac{\sqrt{9+3 {x_i}_0^2+{x_i}_0^4}}{\sqrt{9+3 x_i^2+x_i^4}}
\end{equation}
So, the potential will be now
 \begin{equation}
t_{\pi \Delta (d) \rightarrow \pi \Delta (d)} = -\frac{\gamma_{33}}{m_{\pi}^5} q_3^4 B_3^2
\end{equation}
\begin{equation}
t_{\pi \Delta (d) \rightarrow \pi N (d)} = -\frac{\gamma_{34}}{m_{\pi}^5} q_3^2 q_4^2 B_3 B_4
\end{equation}
\begin{equation}
t_{\pi N (d) \rightarrow \pi N (d)} = -\frac{\gamma_{44}}{m_{\pi}^5} q_4^4 B_4^2
\end{equation}
Using this notation we obtain a more convenient framework for the fit.

\section{Fitting the data}
In order to determine the unknown parameters $\gamma_{ij}$ we make a fit of the partial wave amplitude data of the $\pi N$ scattering in $D_{13}$ of Ref.~\cite{arndt06}. In our case we fit both the real and imaginary parts in an energy range of 1400 to 1800 MeV using the gradient fit method of $\chi^2$. In the fit we have not only the parameters of the transition with L=2, but also the subtraction constants $\alpha_i$ of the loop function. We know that with a regularization scale of $\mu=630$, the subtraction constant have a natural size of -2 for an s-wave, but as shown in Ref. \cite{Roca:2006sz} their fit gives subtraction constants for the d-wave larger than in the case of s-wave amplitudes.

For the analysis of the experimental data we need to normalize the amplitude using Eq.~(7) of Ref. \cite{Roca:2006sz} which relates our amplitude with the experimental one.
\begin{equation}
\tilde{T}_{ij}(\sqrt{s})=-\sqrt{\dfrac{M_i q_i}{4 \pi \sqrt{s}}}\sqrt{\dfrac{M_j q_j}{4 \pi \sqrt{s}}} T_{ij}(\sqrt{s})
\end{equation}
where $M$ and $q$ are the baryon mass an the on-shell momentum of the specific channel.

\begin{figure}
\includegraphics[scale=1]{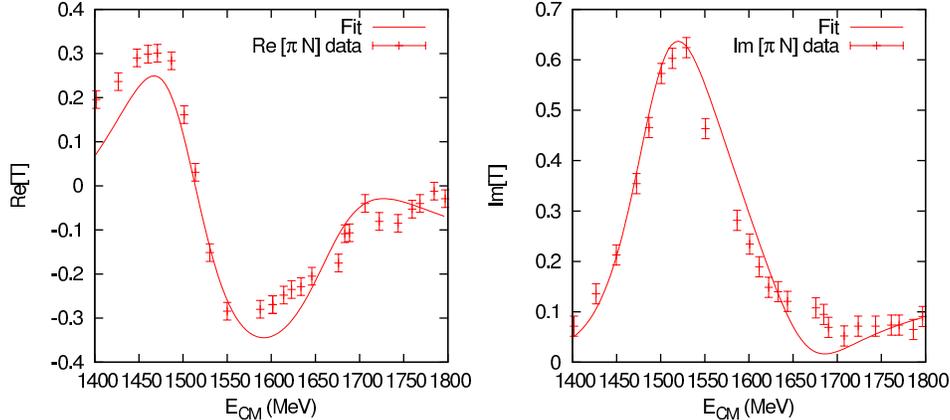}
\caption{Fit to the data of $\pi N$ (d-wave) of Ref.~\cite{arndt06}.}
\label{fig:fit}
\end{figure}

In Fig. \ref{fig:fit} we can see the result of the fit, and in Table \ref{tab:fit} we show the parameters obtained in the fit.
For the estimation of the theoretical errors we follow the criteria of Ref. \cite{Roca:2006sz} where they modify the value of the parameter until the $\chi^2$ increases eight units (which is the procedure to get a 68$\%$ confidence level in the case of seven parameters). 

\begin{table}
\begin{tabular}{ccccccc}
\hline
\hline
$\alpha_{\rho N (s)}$&$\alpha_{\pi \Delta (s)}$&$\alpha_{\pi \Delta (d)}$&$\alpha_{\pi N (d)}$&
$\gamma_{33}$&$\gamma_{34}$&$\gamma_{44}$ \\
\hline
-1.57&-1.6&-4.2&-4.2&0.041&-0.0202&0.0050 \\
~0.03&~0.4&~0.7&~0.3&0.003&~0.0006&0.0003 \\
\hline
\hline
\end{tabular}
\caption{Results of the parameters obtained with the fit. The first row are the parameters and the second row their errors.}
\label{tab:fit}
\end{table}

\section{Results}

\begin{figure}
\includegraphics[scale=1]{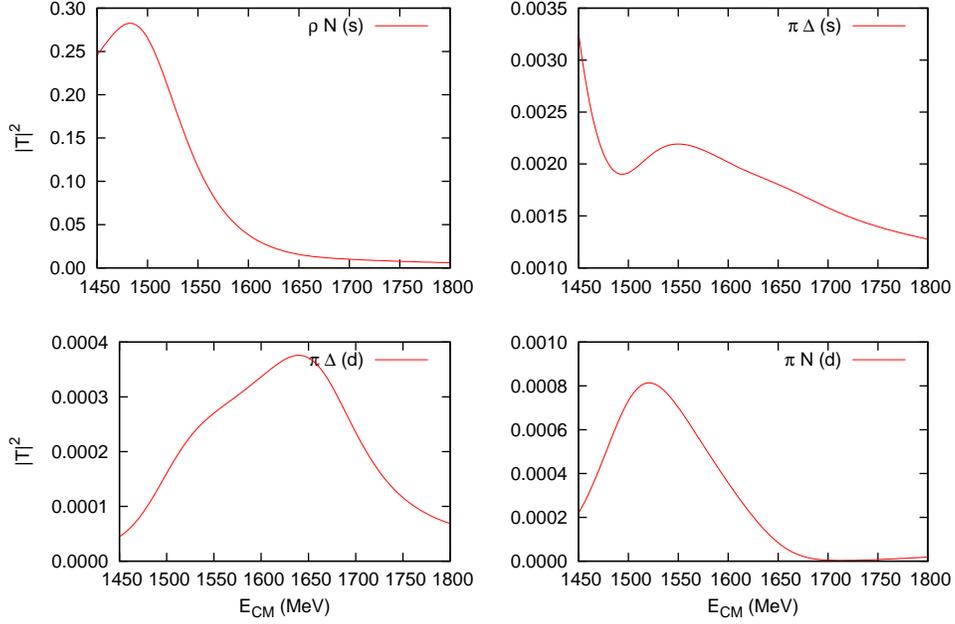}
\caption{Results for the $|T|^2$ matrix.}
\label{fig:t2res}
\end{figure}

\begin{figure}
\includegraphics[scale=1]{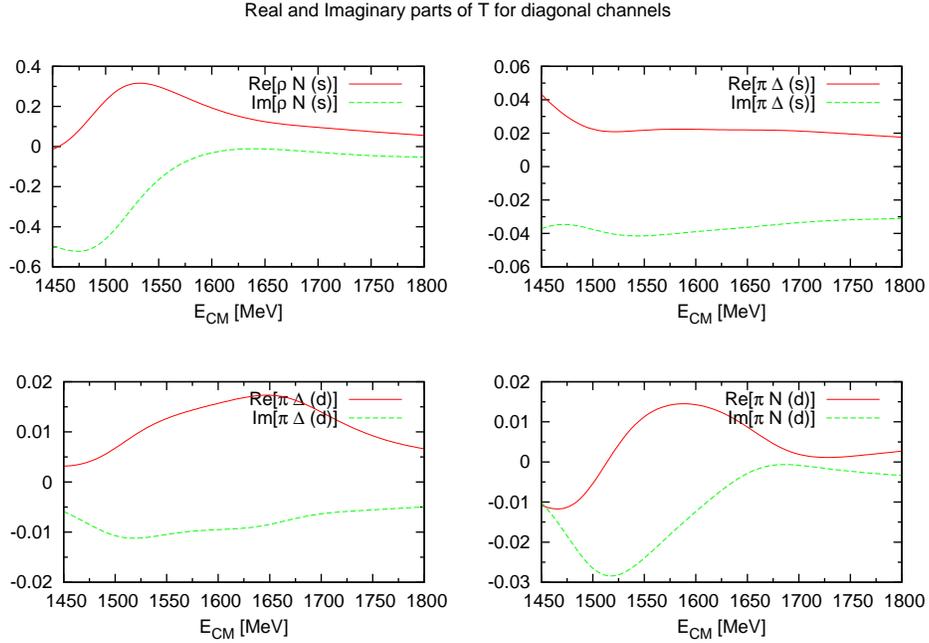}
\caption{Results of real (solid) and imaginary (dashed) parts of T for the diagonal channels.}
\label{fig:treim}
\end{figure}

Using the potentials for the transitions shown in the previous section, we construct the scattering t-matrix using the Bethe-Salpeter equation. The results for $|T|^2$ are shown in Fig. \ref{fig:t2res} for each diagonal transition.
We also include the results of the real and imaginary parts of the t-matrix of the diagonal channels in Fig. \ref{fig:treim}.

In the analysis of the t-matrix we found two poles that can be associated to the resonances $N^*(1520)$ and $N^*(1700)$ respectively. The poles are found in the second Riemann sheet as explained in section II, and the couplings are obtained through the residues of the poles. These results are compiled in Table \ref{tab:couplings}.
Using the couplings and Eq.~(\ref{eq:decaywidth}) one can determine the partial decay widths of the states to each channel. We show these results in Tables \ref{tab:widths1exp} and \ref{tab:widths2exp}, and the experimental results of the PDG \cite{pdg} as well.
As the PDG average has big uncertainties, we consider appropriate to include also some single results of the experiments and analysis \cite{Manley:1992yb,Cutkosky:1979fy, Vrana:1999nt, Thoma:2007bm}.

\begin{table}
\begin{tabular}{c|cc|cc}
\hline
 & \multicolumn{2}{c|}{$N^*(1520) D_{13}$}  & \multicolumn{2}{c}{$N^*(1700) D_{13}$} \\
\hline
Pole & \multicolumn{2}{c|}{1467+i83}  & \multicolumn{2}{c}{1665+i78} \\
\hline
Channel & $g_i$ & $|g_i|$ & $g_i$ & $|g_i|$\\
\hline
$\rho N (s)$ 
&~6.18-1.63i & 6.39
&~1.49+0.42i & 1.55  \\
$\pi \Delta (s)$
&~0.88+0.76i & 1.14    
&-0.39+0.12i & 0.41\\
$\pi \Delta (d)$
&-0.75-0.14i & 0.77      
&~0.50-0.50i & 0.70\\
$\pi N (d)$
&-1.51-0.51i & 1.60    
&-0.09-0.94i & 0.94\\
\hline
\end{tabular}
\caption{Couplings of the resonances for each channel.}
\label{tab:couplings}
\end{table}

\begin{table}
\begin{tabular}{llllcccc}
\hline
\hline
\multicolumn{8}{c}{$N^*(1520) D_{13}$}  \\
\hline	
&&\multicolumn{2}{c}{Briet-Wigner}&\multicolumn{4}{c}{Branching ratio ($\Gamma_i / \Gamma(\%)$)}\\
&Pole&Mass(MeV)&$\Gamma$(MeV)&$\rho N_{(s)}$&$\pi \Delta_{(s)}$&$\pi \Delta_{(d)}$&$\pi N_{(d)}$	\\
\hline
This work	& (1467,83)  & 				&			&5.09&18.37		&8.25		& 65.38		\\
PDG \cite{pdg}		& (1510,55)  & 1515-1525 &100-125 	& 15-25 	&  10-20 	& 10-15 	& 55-65		\\
Manley92 \cite{Manley:1992yb}	 &  		 & 1524$\pm$4	&124$\pm$8	&21$\pm$4&~5$\pm$3&15$\pm$4&59$\pm$3\\
Manley12 \cite{Shrestha:2012ep}	 & (1501,56) & 1512.6$\pm$0.5&117$\pm$1 &20.9$\pm$0.7 &9.3$\pm$0.7&6.3$\pm$0.5&62.7$\pm$0.5\\
Cutkosky79 \cite{Cutkosky:1979fy}& (1510,57) & 1525$\pm$15 	&125$\pm$25	&&&&58$\pm$3\\ 
Vrana00 \cite{Vrana:1999nt} 	& (1504,56)  & 1518$\pm$3	&125$\pm$4 	&~9$\pm$1&15$\pm$2&11$\pm$2&63$\pm$2\\
Toma08 \cite{Thoma:2007bm} 		& (1509,57)	 & 1520$\pm$10  &125$\pm$15 &13$\pm$5&12$\pm$4&14$\pm$5&58$\pm$8\\
Anisovich12 \cite{thoma}		& (1507,56)  & 1517$\pm$3	&114$\pm$4  &&19$\pm$4& 9$\pm$2&62$\pm$3\\
Ardnt06	\cite{arndt06}			& (1515,57)  & 1514.5$\pm$0.2&103.6$\pm$0.4&&&&63.2$\pm$0.1\\
\hline
\hline
\end{tabular}
\caption{Results of the partial decay widths for the $N^*(1520)$ resonance.}
\label{tab:widths1exp}
\end{table}

\begin{table}
\begin{tabular}{llllcccc}
\hline
\hline
\multicolumn{8}{c}{$N^*(1700) D_{13}$}  \\
\hline	
&&\multicolumn{2}{c}{Briet-Wigner}&\multicolumn{4}{c}{Branching ratio ($\Gamma_i / \Gamma(\%)$)}\\
&Pole&Mass(MeV)&$\Gamma$(MeV)&$\rho N_{(s)}$&$\pi \Delta_{(s)}$&$\pi \Delta_{(d)}$&$\pi N_{(d)}$	\\
\hline
This work	& (1665,78)  		 & 			 &			&8.02&4.51&13.36&28.54	\\
PDG \cite{pdg}					 & (1700,75) & 1650-1750 &100-250 	& $<$35 &  10-90 & $<$20 & 12$\pm$5		\\
Manley92 \cite{Manley:1992yb}	 &  		 & 1737$\pm$44	&249$\pm$218&13$\pm$17&5$\pm$10&80$\pm$19&1$\pm$2\\
Manley12 \cite{Shrestha:2012ep}	 & (1662,55) & 1665$\pm$3   &56$\pm$8   &38$\pm$6 &31$\pm$9&3$\pm$2&2.8$\pm$0.5\\
Cutkosky79 \cite{Cutkosky:1979fy}& (1660,38) & 1670$\pm$25 	&80$\pm$40	&&&&11$\pm$5\\ 
Vrana00 \cite{Vrana:1999nt} 	& (1704,78)  & 1736$\pm$33	&175$\pm$133 	&7$\pm$1&11$\pm$1&79$\pm$56&4$\pm$1\\
Toma08 \cite{Thoma:2007bm} 		& (1710,78)	 & 1740$\pm$20  &180$\pm$30 &20$\pm$15&10$\pm$5&20$\pm$11&$8^{+8}_{-4}$\\
Anisovich12 \cite{thoma}		& (1770,210) & 1790$\pm$40	&390$\pm$140  &&72$\pm$23& $\leq$10&12$\pm$5\\
\hline
\hline
\end{tabular}
\caption{Results of the partial decay widths for the $N^*(1700)$ resonance.}
\label{tab:widths2exp}
\end{table}

As we can see in Table \ref{tab:couplings}, the $N^*(1520)$ couples mostly to the channel $\rho N$, which is closed for the nominal mass of the $\rho$, but although the mass of the resonance is under the $\rho N$ threshold, using Eq.~(\ref{prhocon}) we can generate a momentum giving a small partial decay width. In comparison with the experiments, the decay width of $\rho N$ to $N^*(1520)$ is smaller but of the same order of magnitude. Note that in experimental analyses one evaluates this rate subtracting the other ones from the total width. In either method the uncertainties for this closed channel are necessarily large. In the case of $N^*(1700)$ the coupling of $\rho N$ is smaller but, as we are closer to the threshold, the decay width is of the same order of magnitude as for the $N^*(1520)$. The result for the decay to $\rho N$ is in a good agreement with experiment considering the experimental uncertainties.

The $\pi \Delta$ (s-wave) channel has small couplings for $N^*(1700)$ but, as we are over the threshold, the phase space is big enough to generate a moderate decay width. The results of the branching ratios are in a fair agreement with the experiments within the large uncertainties. For this case the PDG average has a very wide range, but the individual results are more precise. There is a large disagreement with the result of \cite{thoma} but it is much closer to the one of \cite{Thoma:2007bm} by the same group.

The $\pi \Delta$ (d-wave) channel has the lower coupling to $N^*(1520)$ but, since it has a large momentum, the partial decay width is bigger than for the $\rho N$ channel. The branching ratio for $N^*(1520)$ agrees with experimental results, but in the case of $N^*(1700)$, although the result is compatible with some experimental branching ratios, these results are very different and in some cases have large errors.

Finally, for the channel $\pi N$ (d-wave), since we fit the amplitude to the data, the result of the branching ratio of $N^*(1520)$ is in a very good agreement with the experimental values. 
On the other hand, we get a branching ratio of the $N^*(1700)$ for the decay into $\pi\Delta$ (d-wave) which is in line with experimental determinations but about twice as large as the PDG average. Yet, an inspection to Table \ref{tab:widths2exp} indicates that the dispersion of experimental data for the $N^*(1700)$ is quite large, for what further attention to this resonance should be most welcome.

It is interesting to compare the present results with those of the J\"ulich group \cite{Doring:2009yv,Ronchen:2012eg}. In those works the authors also consider $\rho N$, $\pi N$ and $\pi\Delta$ channels, in s or d-waves when allowed for $J^P=3/2^-,I=1/2$. The dynamics used in those works in similar to the one used here up to details concerning a contact $\rho N$ term, present in \cite{Ronchen:2012eg}  (see Figs. 2 and 3), the Kroll-Ruderman term for $\rho N \rightarrow \pi\Delta(s)$ no considered in \cite{Doring:2009yv,Ronchen:2012eg} and the use of form factors in those work while we use dimensional regularization. In all cases we adhere to the dynamics of local hidden gauge approach. One novelty in our work is that the $N^*(1520)$ is dynamically generated in our approach, while in \cite{Doring:2009yv,Ronchen:2012eg} is an input resonance. The $N^*(1520)$ was obtained as a dynamically generated state in \cite{Sarkar:2009kx} using coupled channels and input from chiral lagrangians. In the present case we have taken only the main $\pi \Delta (s)$ channel, but it is well known that effects of other channels can be incorporated by small changes in the subtraction constants of dimensional regularization, which in the present problem we have left as free parameters. 

The other novelty is the $N^*(1700)$ that we also obtain as dynamically generated with our input. It is interesting to remark that in the work of \cite{Doring:2009yv,Ronchen:2012eg}, with the input of non pole terms there was also a state around 1700 MeV which was dynamically generated and a pole could be seen in the third Riemann sheet defined there. However when the pole terms were added, the pole in the amplitude fades away in that approach. In our work the pole appears around 1665 MeV together with the one for the $N^*(1520)$. The unitarization of the driving terms in all the channels that we consider, including the transition between all of them, generates in our case the two poles and their interference is responsible for the vanishing amplitude of $\pi N$(d) around 1700 MeV, which is consistent with the experiment. However, physical traces experimentally identifiable as a peak, remain in the $\pi\Delta$(d) channel as one can see in Fig.\ref{fig:t2res}, hence, the more intuitive picture of a peak in a cross section is also seen here but only in the $\pi N$(d) channel.
Actually in \cite{Manley:1992yb,Manley:1984jz}, the most characteristic feature attributed to the $N^*(1700)$ was its large coupling to $\pi N$(d).

We should also mention that there are other works that mix pseudoscalars and vectors \cite{kanchan1,kanchan2,kanchan3} and others that also include decuplet of baryons \cite{juan}, this latter one invoking SU(3) spin-isospin symmetry. In our approach we have followed strictly the local hidden gauge formalism that has proved to be very successful in a variety of processes (see recent review \cite{review}).
 
\section{Conclusions}

We have done a theoretical study for the meson-baryon scattering in the region of $\sqrt{s}=1400-1800~MeV$ with $J^P=3/2^-$. We considered the standard coupled channels used in the most complete experimental analyses, $\rho N$ (s-wave), $\pi \Delta$ (s-wave), $\pi \Delta$ (d-wave) and $\pi N$ (d-wave). The interaction of these channels was taken from the local hidden gauge approach and the loops were regularized using dimensional regularization with subtraction constant of natural size. These constants were varied within a moderate range to obtain a good fit to the $\pi N$ (d-wave) data. After this, the rest are predictions of the theory.

The first important theoretical finding is that the model obtained, after fitting exclusively the $\pi N$ (d-wave) data, produces two poles: one around 1480 MeV that we associate to the $N^*(1520) (3/2^-)$ and another around 1670 MeV that we associate to the $N^*(1700) (3/2^-)$. It is worth noting that the presence of the two poles is rather solid, since they remain by making changes in the parameters that do not spoil massively the agreement with the $\pi N$ (d-wave) experimental data.

With the model obtained we determined partial decay widths to all the channels. We found an excellent agreement with experiments for the data on the $N^*(1520) (3/2^-)$ and rough to fair for the $N^*(1700) (3/2^-)$. Yet, we noticed the large dispersions of experimental data for the $N^*(1700) (3/2^-)$. The study done here gives a boost to the existence of the $N^*(1700)$ which has been questioned in some recent experimental analyses. In view of this extra support for the $N^*(1700)$ and the large dispersion of the data, further experimental studies concentrating in this energy region for the quantum numbers $(3/2^-)$ of this resonance should be encouraged.

\section*{Acknowledgments}

We would like to thank M. D\"oring, D. M. Manley, I. I. Strakovsky, D. Bugg and R. L. Workman for a through reading of our paper and valuable comments.
This work is partly supported by the Spanish Ministerio de Economia y Competitividad and European FEDER funds under the contract number FIS2011-28853-C02-01, and the Generalitat Valenciana in the program Prometeo, 2009/090. We acknowledge the support of the European Community-Research Infrastructure Integrating Activity Study of Strongly Interacting Matter (acronym HadronPhysics3, Grant Agreement n. 283286) under the Seventh Framework Programme of EU.
This work is also partly supported by the National Natural Science Foundation of China under grant  11105126.

\appendix
\section{Evaluation the vertices.}
\label{app:vertices}
\begin{figure}
\includegraphics[width=0.2\textwidth]{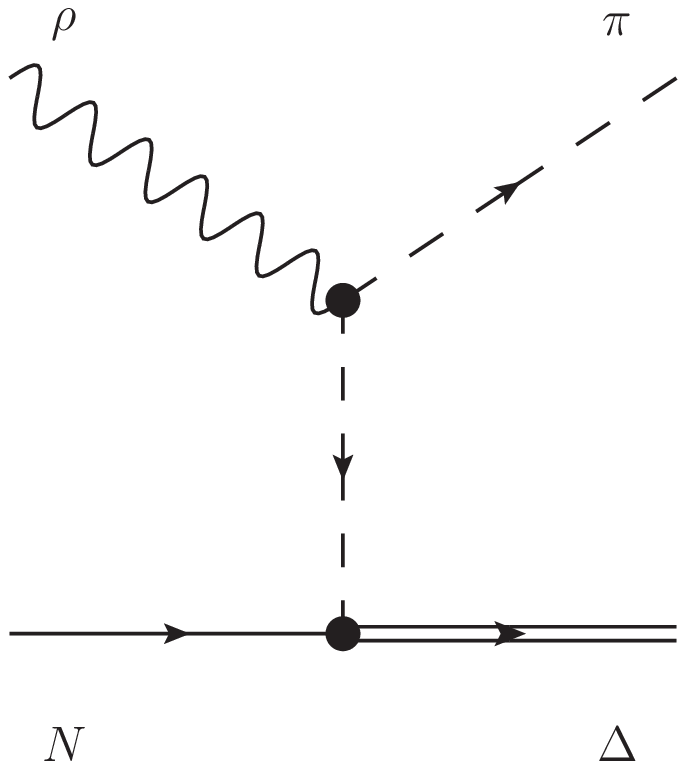}
\caption{Diagram of the transition $\rho N \rightarrow \pi \Delta$ (s-wave)}
\label{fig:rhonpidel}
\end{figure}

\begin{figure}
\includegraphics[width=0.5\textwidth]{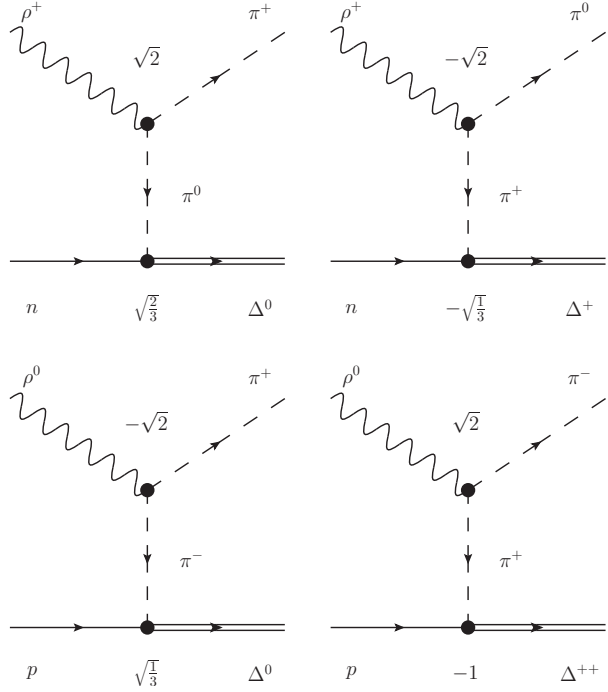}
\caption{Diagram of the transition $\rho N \rightarrow \pi \Delta$ (s-wave)}
\label{fig:rhonpidelall}
\end{figure}

The diagonal transition for $\pi \Delta$ (s-wave) and $\rho N$ (s-wave) are taken from the references \cite{Sarkar:2009kx} and \cite{Oset:2009vf} respectively.
So we need to evaluate the transition potential of Fig. \ref{fig:rhonpidel}.
When constructing the Kroll-Ruderman term, we shall get the contact term of the type of $\vec{S}^+ \vec{\epsilon}$. We must evaluate this matrix element
\begin{equation}
\left\langle 3/2 M^\prime | \vec{S}^+ \vec{\epsilon} | J M \right\rangle
\end{equation}
where
\begin{equation}
\left. | J M \right\rangle = \sum_m {\cal C} \left( 1/2, 1, J; m, M-m,M\right) \left. |1/2 m\right\rangle \left. | \vec{\epsilon}_{M-m} \right\rangle
\end{equation}
As $\vec{\epsilon}$ is part of $\left. | J M \right\rangle$, we get that in the spherical basis
\begin{equation}
\vec{S}^+ \vec{\epsilon}_{M-m} \equiv S^+_{M-m} 
\end{equation}
then the matrix element becomes
\begin{equation}
\left\langle 3/2 M^\prime | S^+_{M-m} | 1/2 m \right\rangle \equiv {\cal C} \left(1, 1/2, 3/2; M-m,m, M^\prime \right) \left\langle || S^+ || \right\rangle
\end{equation}
where the reduced matrix element is chosen to be 1 by construction, and from the Clebsch-Gordan coefficient we get a $\delta_{M M^\prime}$. Finally we get that
\begin{equation}
\sum_m {\cal C} \left( 1/2, 1, J; m, M-m,M\right) {\cal C} \left( 1/2, 1, 3/2; m, M-m,M\right) \delta_{M M^\prime} = \delta_{M M^\prime} \delta_{3/2 J}
\end{equation}
Hence, for spin $3/2$ of $\rho N$ this operator is unity 1. Getting the Kroll-Ruderman term, we must substitute
\begin{equation}
\vec{\epsilon} \left(\vec{P}_V+2 \vec{q} \right) \frac{1}{(P_V+q)^2-m_\pi^2}\vec{S}^+ \left(\vec{P}_V+ \vec{q} \right) \rightarrow \vec{S}^+ \vec{\epsilon}
\end{equation}
We need to write the states of isospin basis in charge basis, with our sign convention $\rho^+\equiv-\left.|1,1\right\rangle$.
\begin{eqnarray}
\left.|\rho N, I=1/2,1/2 \right\rangle &=& -\sqrt{\frac{2}{3}} \left. | \rho^+ n \right\rangle - \frac{1}{\sqrt{3}} \left. | \rho^0 p \right\rangle \\
\left.|\rho N, I=3/2,1/2 \right\rangle &=& -\sqrt{\frac{1}{3}} \left. | \rho^+ n \right\rangle + \sqrt{\frac{2}{3}} \left. | \rho^0 p \right\rangle \\
\left.|\pi \Delta, I=1/2,1/2 \right\rangle &=& \sqrt{\frac{1}{2}} \left. | \pi^- \Delta^{++} \right\rangle - \sqrt{\frac{1}{3}} \left. | \pi^0 \Delta^{+} \right\rangle - \sqrt{\frac{1}{6}} \left. | \pi^+ \Delta^{0} \right\rangle\\
\left.|\pi \Delta, I=3/2,1/2 \right\rangle &=& -\sqrt{\frac{2}{5}} \left. | \pi^- \Delta^{++} \right\rangle - \sqrt{\frac{1}{15}} \left. | \pi^0 \Delta^{+} \right\rangle - \sqrt{\frac{8}{15}} \left. | \pi^+ \Delta^{0} \right\rangle\\
\end{eqnarray}
For the transition $\rho N \rightarrow \pi \Delta$ (s-wave) we have the vertices of Fig. \ref{fig:rhonpidelall}. For evaluating those vertices we need both lagrangians, one for the VPP vertex and the other one for the lower $\pi N \Delta$ vertex.
The first diagram has a vertex
\begin{equation}
t_{\rho^+ \pi^+ \pi^0} = - {\cal L} = g \sqrt{2} \left( \vec{P}_V +\vec{q}+\vec{q} \right) \vec{\epsilon}
\end{equation}
The other coefficients of the diagrams for the different charge combinations are indicated in Fig. \ref{fig:rhonpidelall}. The lower vertex has an isospin coefficient for $\pi N \Delta$, giving a transition vertex for the first diagram
\begin{equation}
-i t_{\pi^0 n \Delta^0} = \frac{f_{\pi N \Delta}}{m_\pi} \vec{S}^+ \left( \vec{P}_V+\vec{q} \right) \sqrt{\frac{2}{3}}
\end{equation}
The other isospin coefficients of the lower vertices are shown in Fig. \ref{fig:rhonpidelall}.
Combining the two vertices and the pion propagator, we get the transition potential of the first diagram.
\begin{equation}
-i t_{\rho^+ n \rightarrow \pi^+ \Delta^0} = -i g \sqrt{2} 2\vec{q} \vec{\epsilon} \frac{i}{(P_V+q)^2-m_\pi^2} \frac{f_{\pi N \Delta}}{m_\pi} \vec{S}^+ \vec{q} \sqrt{\frac{2}{3}}
\end{equation}
As mentioned before, we need to add the Kroll-Ruderman term as $\vec{S}^+ \vec{\epsilon}$,
\begin{equation}
-i t_{\rho^+ n \rightarrow \pi^+ \Delta^0 (KR)} = g \sqrt{2} \frac{f_{\pi N \Delta}}{m_\pi} \sqrt{\frac{2}{3}} \vec{S}^+ \vec{\epsilon} 
\end{equation}
So in the s-wave for a $\pi$-exchange we can sum the momenta products as
\begin{equation}
2\vec{q}\cdot\vec{\epsilon}~\vec{S}^+\cdot\vec{q} =2 q_i \epsilon_i S^+_j q_j \rightarrow \epsilon_i  S^+_j \frac{2}{3} \vec{q}~^2 \delta_{ij} = \frac{2}{3} \vec{q}~^2 \vec{S}^+\cdot \vec{\epsilon}
\label{eq:swaveq2}
\end{equation}
And we have shown for $J=3/2$ that $\vec{S}^+\cdot \vec{\epsilon}=1$
The sum of the pion propagator and the Kroll-Ruderman for the first diagram gives us
\begin{equation}
-i t_{\rho^+ n \rightarrow \pi^+ \Delta^0} = g \sqrt{2} \sqrt{\frac{2}{3}} \frac{f_{\pi N \Delta}}{m_\pi} \left( \frac{\frac{2}{3} \vec{q}~^2}{(P_V+q)^2-m_\pi^2}+1 \right) 
\end{equation}

Hence putting all the coefficients of the Clebsch-Gordan and the corresponding vertex factor we get
\begin{eqnarray}
&&\left\langle \pi \Delta, I=1/2,1/2 | - i t | \rho N, I=1/2,1/2 \right\rangle =\\
&&\langle \sqrt{\frac{1}{2}} \pi^- \Delta^{++} - \sqrt{\frac{1}{3}} \pi^0 \Delta^{+} - \sqrt{\frac{1}{6}} \pi^+ \Delta^{0} |-it| \sqrt{\frac{2}{3}} \rho^+ n - \frac{1}{\sqrt{3}} \rho^0 p \rangle = \\
&&\left( 
 \frac{1}{\sqrt{2}} \frac{1}{\sqrt{3}} \sqrt{2} (-1)
-\frac{1}{\sqrt{3}} \sqrt{\frac{2}{3}} (-\sqrt{2}) (-\frac{1}{\sqrt{3}})
-\frac{1}{\sqrt{6}} \sqrt{\frac{2}{3}} \sqrt{2} \sqrt{\frac{2}{3}}
-\frac{1}{\sqrt{6}} \frac{1}{\sqrt{3}} (-\sqrt{2}) \frac{1}{\sqrt{3}} \right) = \frac{2}{\sqrt{3}} \\
\end{eqnarray}
The isospin factor is $\frac{2}{\sqrt{3}}$, so finally all together
\begin{equation}
-i t_{\rho N (s) \rightarrow \pi \Delta (s)} = g \frac{2}{\sqrt{3}} \frac{f_{\pi N \Delta}}{m_\pi} \left( \frac{\frac{2}{3} \vec{q}~^2}{(P_V+q)^2-m_\pi^2}+1 \right) 
\end{equation}

We need also to evaluate the s-wave and d-wave mixing, $\rho N$ is in $L=0$ but $\pi \Delta$ can be in $L=0$ and $L=2$ since there is enough momentum for $N^*(1520)$ and $N^*(1700)$. With $\rho N$ below threshold there is no need to worry about $L=2$ for $\rho N$. We want to work with $\rho N$(s), $\pi \Delta$(s) and $\pi \Delta$(d), for this we shall also include  the $\rho N \rightarrow \pi N (d)$ transition.

\begin{figure}
\includegraphics[width=0.7\textwidth]{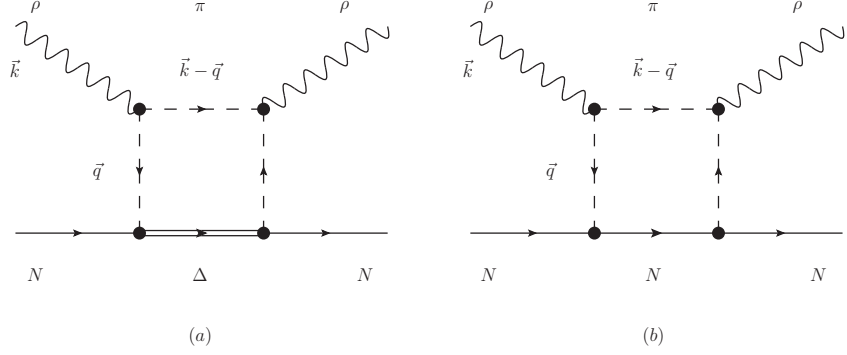}
\caption{Box diagrams for the $L=2$ transition: (a) $\rho N \rightarrow \pi \Delta$ and (b) $\rho N \rightarrow \pi N$ }
\label{fig:rhonboxs}
\end{figure}

Let us evaluate the transition $\rho N (s) \rightarrow \rho N (s)$ thought an intermediate state $\pi \Delta$ as shown in Fig~\ref{fig:rhonboxs}(a). Let us take the third component of spin $1/2$ and $1$ for $N$ and $\rho$ respectively, to get $M=3/2$ since the result does not depend on the third component. With this choice, both polarizations $\vec{\epsilon}$ and $\vec{\epsilon}~^\prime$ will be the same and we have neglected the momentum $\vec{k}$. The terms of the four vertex are
\begin{equation}
\vec{\epsilon}\cdot 2\vec{q}~\vec{S}\cdot \vec{q}~\vec{\epsilon}\cdot2\vec{q}~\vec{S}^+\cdot\vec{q}
\end{equation}
We can put together the polarization terms and the isospin transition operators and use that
\begin{equation}
\vec{S}\cdot\vec{q}~\vec{S}^+\cdot\vec{q}=S_i q_j~S^+_j q_j = (\frac{2}{3}\delta_{ij}-\frac{i}{3}\varepsilon_{ijk} \sigma_k)q_i q_j=\frac{2}{3}\vec{q}~^2
\end{equation}
For the polarization terms we have in the loop integral
\begin{equation}
4 \vec{\epsilon}\cdot\vec{q}~\vec{\epsilon}\cdot\vec{q}=4 \epsilon_i q_i~\epsilon_j q_j = \frac{4}{3} \delta_{ij}  \vec{q}~^2 \epsilon_i \epsilon_j= \frac{4}{3} \vec{q}~^2 \vec{\epsilon} \cdot \vec{\epsilon} = \frac{4}{3}  \vec{q}~^2
\label{eq:pola}
\end{equation}
as $\vec{\epsilon} \cdot \vec{\epsilon} = 1$. Now, all together we have
\begin{equation}
\frac{4}{3}  \vec{q}~^2 \frac{2}{3}  \vec{q}~^2
\label{eq:totalds}
\end{equation}
including both $L=0$ and $L=2$ contributions. For s-wave we got $\frac{2}{3} \vec{q}~^2$ for one pion exchange (See Eq.(\ref{eq:swaveq2})) so, the box equivalent operator with only s-wave will be
\begin{equation}
\frac{2}{3} \frac{2}{3}  \vec{q}~^2   \vec{q}~^2 = \frac{4}{9} \vec{q}~^4
\end{equation}
subtracting this from the total transition (Eq.~(\ref{eq:totalds})) we get the d-wave transition in the box
\begin{equation}
\frac{8}{9}  \vec{q}~^4 - \frac{4}{9} \vec{q}~^4 = \frac{4}{9} \vec{q}~^4
\end{equation}
So finally we obtain that for the d-wave transition of one pion exchange the contribution is $\frac{2}{3} \vec{q}~^2$, the same one as for the s-wave transition. All the other terms are the same but without the Kroll-Ruderman factor which only comes in $L=0$. Thus,
\begin{equation}
-i t_{\rho N (s) \rightarrow \pi \Delta (d)} = g \frac{2}{\sqrt{3}} \frac{f_{\pi N \Delta}}{m_\pi} \left( \frac{\frac{2}{3} \vec{q}~^2}{(P_V+q)^2-m_\pi^2}\right) 
\end{equation}
In the case of $\rho N \rightarrow \pi N$ (d-wave) transition, the s-wave for $3/2$ does not exist, so whatever comes out will be d-wave. We use the same procedure as for $\pi \Delta$, we chose the third component of spin to be $3/2$. The diagram is shown in Fig~\ref{fig:rhonboxs}(b) and the operators of the vertices are
\begin{equation}
\vec{\epsilon}\cdot 2\vec{q}~\vec{\sigma}\cdot \vec{q}~\vec{\epsilon}\cdot2\vec{q}~\vec{\sigma}\cdot\vec{q}
\end{equation}
In the same way
\begin{equation}
\vec{\sigma}\cdot\vec{q}~\vec{\sigma}\cdot\vec{q}=\sigma_i q_j~\sigma_j q_j = (\delta_{ij}-i\varepsilon_{ijk} \sigma_k)q_i q_j=\vec{q}~^2
\end{equation}
and for the polarization we have the same result as Eq.~(\ref{eq:pola}). So we get
\begin{equation}
\frac{4}{3}  \vec{q}~^2 \vec{q}~^2 = \frac{2}{\sqrt{3}} \vec{q}~^2 \frac{2}{\sqrt{3}} \vec{q}~^2
\end{equation}
The result for each vertex in the $\rho N \rightarrow \pi N$ (d-wave) is $\frac{2}{\sqrt{3}} \vec{q}~^2$.

Hence the $\vec{\epsilon}\cdot 2\vec{q}~\vec{\sigma}\cdot\vec{q}$ operator in one pion exchange can be replaced by
\begin{equation}
\frac{2}{\sqrt{3}}\vec{q}~^2 \frac{f_{\pi N N}}{m_\pi}
\end{equation}
which replaces the equivalent contribution $\frac{2}{3} \vec{q}~^2 \frac{f_{\pi N \Delta}}{m_\pi}$ that we had for the $\rho N \rightarrow \pi \Delta$ (d-wave) transition.

\begin{figure}
\includegraphics[width=0.5\textwidth]{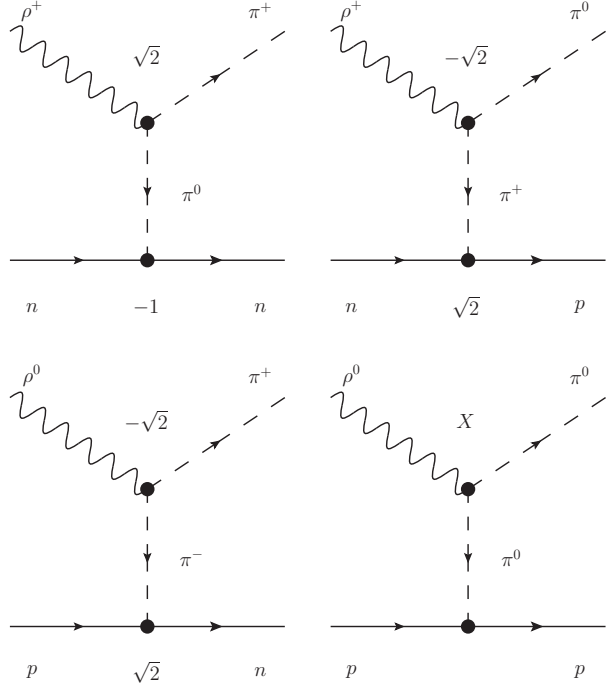}
\caption{Diagram of the transition $\rho N \rightarrow \pi N$ (d-wave)}
\label{fig:rhonpinall}
\end{figure}
Next we must do the isospin combination and the charge factors for each vertex as shown in Fig.~\ref{fig:rhonpinall}.
\begin{eqnarray}
\left.|\rho N, I=1/2,1/2 \right\rangle &=& -\sqrt{\frac{2}{3}} \left. | \rho^+ n \right\rangle - \frac{1}{\sqrt{3}} \left. | \rho^0 p \right\rangle \\
\left.|\pi N, I=1/2,1/2 \right\rangle &=& -\sqrt{\frac{2}{3}} \left. | \pi^+ n \right\rangle - \frac{1}{\sqrt{3}} \left. | \pi^0 p \right\rangle
\end{eqnarray}
Looking at the diagrams, the isospin factor for the transition $\rho N \rightarrow \pi N$ (d-wave) will be
\begin{eqnarray}
&&\left\langle \rho N, I=1/2,1/2 | T | \pi N, I=1/2,1/2 \right\rangle = \\
&&\frac{2}{3} \left\langle \rho^+ n | \pi^+ n \right\rangle
+\sqrt{\frac{2}{3}} \sqrt{\frac{1}{3}} \left\langle \rho^+ n | \pi^0 p \right\rangle
+\sqrt{\frac{2}{3}} \sqrt{\frac{1}{3}} \left\langle \rho^0 p | \pi^+ n \right\rangle = \\
&&\frac{2}{3} \sqrt{2}(-1)+2\sqrt{\frac{2}{3}} \sqrt{\frac{1}{3}}(-\sqrt{2}) \sqrt{2}=-2\sqrt{2}
\end{eqnarray}
Finally we have
\begin{equation}
-i t_{\rho N (s) \rightarrow \pi N (d)} = g (-2 \sqrt{6}) \frac{f_{\pi N N}}{m_\pi} \left( \frac{\frac{2}{3} \vec{q}~^2}{(P_V+q)^2-m_\pi^2}\right) 
\end{equation}

\end{document}